\documentclass[aps,prd,twocolumn,superscriptaddress,longbibliography,nofootinbib,10pt]{revtex4-2}
\usepackage{graphicx,longtable,mathrsfs,array}
\usepackage[hidelinks]{hyperref}
\usepackage{amsmath}
\usepackage[dvipsnames]{xcolor}
\usepackage{mathrsfs}
\usepackage{amssymb}
\usepackage{orcidlink}
\usepackage{comment}
\usepackage{tabularx}
\usepackage[normalem]{ulem}

\newcommand{\be}{\begin{equation}}
\newcommand{\ee}{\end{equation}}
\newcommand{\bea}{\begin{eqnarray}}
\newcommand{\eea}{\end{eqnarray}}
\newcommand{\Beq}{\begin{equation}\begin{aligned}}
\newcommand{\Eeq}{\end{aligned}\end{equation}}
\newcommand{\Fig}[1]{Fig.~\ref{#1}}
\newcommand{\Eq}[1]{Eq.~(\ref{#1})}
\newcommand{\Eqs}[2]{Eqs.~(\ref{#1}) and (\ref{#2})}
\newcommand{\Sec}[1]{Sect.~\ref{#1}}

\newcommand{\App}[1]{App.~\ref{#1}}

\newcommand{\mpl}{m_{\rm pl}}

\usepackage{color}
\usepackage{ifthen}
\newboolean{editorial}
\setboolean{editorial}{true}
\newcommand{\editorial}[2]{\ifthenelse{\boolean{editorial}}{\textcolor{red}{[\textsf{\textbf{{#1}}}: }\textcolor{blue}{\textsf{{#2}}}\textcolor{red}{]}}{}}

\usepackage{xcolor}

\begin{document}

\title{Primordial Black Holes in a Radiation-Dominated Universe}

\author{Thomas W.~Baumgarte\orcidlink{0000-0002-6316-602X}}
\email{tbaumgar@bowdoin.edu}
\affiliation{Department of Physics and Astronomy, Bowdoin College, Brunswick, ME 04011, USA}

\author{Katy~Clough\orcidlink{0000-0001-8841-1522}}
\email{k.clough@qmul.ac.uk}
\affiliation{Centre for Geometry, Analysis and Gravitation, School of Mathematical Sciences, Queen Mary University of London, Mile End Road, London E1 4NS, United Kingdom}

\author{Mary~Gerhardinger\orcidlink{0000-0001-9359-5960}}
\email{maryge@sas.upenn.edu}
\affiliation{Center for Particle Cosmology, Department of Physics and Astronomy,
University of Pennsylvania, Philadelphia, Pennsylvania 19104, USA}

\author{John~T.~Giblin~Jr.\orcidlink{0000-0003-1505-8670}}
\email{giblinj@kenyon.edu}
\affiliation{Department of Physics, Kenyon College, 201 N College Rd, Gambier, OH 43022, USA}
\affiliation{Department of Physics/CERCA/Institute for the Science of Origins, Case Western Reserve University, Cleveland, OH 44106-7079, USA}
\affiliation{Center for Cosmology and AstroParticle Physics (CCAPP) and Department of Physics, The Ohio State University, Columbus, OH 43210, USA}

\author{Amanda Miller\orcidlink{0000-0002-4637-6906}}
\email{miller.10749@osu.edu}
\affiliation{Department of Physics, Kenyon College, 201 N College Rd, Gambier, OH 43022, USA}
\affiliation{Center for Cosmology and AstroParticle Physics (CCAPP) and Department of Physics, The Ohio State University, Columbus, OH 43210, USA}

\begin{abstract}
Primordial fluctuations frozen out during inflation re-enter the cosmological horizon and can collapse, leading to the formation of primordial black holes. We perform simulations of the direct collapse of over-dense regions re-entering the horizon during a radiation-dominated epoch, using full 3+1 general relativistic simulations with the BSSN formalism. 
Building on previous studies, we impose periodic boundary conditions and allow the matter content of the Universe to self-consistently drive its dynamics. 
We analyze the evolution of over-densities in both the collapse and dispersal regimes and find a threshold,  $0.77<\delta_c<0.83$, above which over-densities collapse and form primordial black holes. 
Our findings are consistent with previous analytic predictions as well as numerical studies that use different formalisms and computational approaches, and hence provide independent validation of those results.
\end{abstract}

\maketitle
\section{Introduction}

Primordial black holes (PBHs) may have formed from the collapse of over-dense regions in the early Universe \cite{Zeldovich:1967lct,Hawking:1971ei,Carr:1975qj,Choptuik:1992jv, Evans:1994pj,Niemeyer:1997mt,Niemeyer:1999ak,Green:1999xm,Musco:2012au,deJong:2021bbo,Escriva:2022pnz,deJong:2023gsx,Escriva:2024lmm,Milligan:2025zbu}.  In the standard picture, over-densities are seeded from quantum fluctuations that freeze out during inflation \cite{Hodges:1990bf,Carr:1993aq,Carr:1994ar,Ivanov:1994pa,Randall:1995dj,Garcia-Bellido:1996mdl,Taruya:1998cz,Bassett:2000ha,Yoo:2020lmg}, leading to large-scale curvature fluctuations that later re-enter the cosmological horizon.
The superposition of multiple fluctuations may occasionally lead to large peaks, which can collapse on re-entry to form PBHs.
The question of when these PBHs form, specifically the threshold above which over-densities collapse, has been studied in myriad contexts, for example in radiation-dominated \cite{Shibata:1999zs,Hawke:2002rf,Musco:2004ak,Yoo:2018pda,Uehara:2024yyp}, matter-dominated \cite{deJong:2021bbo,deJong:2023gsx,Ebrahimian:2025syf,Ye:2025wif}, and scalar-field-dominated \cite{Milligan:2025zbu,Padilla:2025bkv} cosmologies and for various initial profiles, including spherical \cite{Escriva:2021aeh}, ellipsoidal \cite{Escriva:2024lmm, Yoo:2024lhp} and spinning over-densities \cite{Mirbabayi:2019uph,Banerjee:2023qya,Mohammadi:2025avz}.
Most studies, including ours, employ initial data generated from spherically symmetric profiles, justified by the claim in peak theory that the largest excursions from homogeneity tend to be spherically symmetric \cite{Bardeen:1985tr, Bloomfield:2015ila}.\footnote{There are several other formation mechanisms for PBHs, including nonlinear processes during inflation \cite{Clesse:2015wea,Inomata:2017okj,Garcia-Bellido:2017mdw,Ezquiaga:2017fvi} or preheating \cite{Adshead:2025gka}, the collision of bubbles of first order phase transitions \cite{Crawford:1982yz,Hawking:1982ga,Kodama:1982sf,Moss:1994iq,Khlopov:1998nm,Khlopov:1999ys,Leach:2000ea,Khlopov:2000js,Kitajima:2020kig,Kawana:2021tde,Musco:2023dak,Ning:2026nfs}, the collapse of cosmic strings \cite{Kibble:1976sj,Hogan:1984zb,Hawking:1987bn,Polnarev:1988dh,Garriga:1993gj,Caldwell:1995fu,MacGibbon:1997pu,Wichoski:1998ev,Hansen:1999su,Nagasawa:2005hv,Bramberger:2015kua,Helfer:2018qgv,James-Turner:2019ssu,Bertone:2019irm,Jenkins:2020ctp,Aurrekoetxea:2020tuw,Jenkins:2020ctp,Blanco-Pillado:2021klh}, the collapse of domain walls from a first order phase transition \cite{Rubin:2000dq,Dokuchaev:2004kr,Liu:2019lul}, the collapse of a scalar condensate \cite{Cotner:2016cvr,Cotner:2017tir}, and some baryogenesis scenarios \cite{Dolgov:1992pu,Dolgov:2008wu,Kannike:2017bxn,Dolgov:2020sov}.  
Here, however, we focus solely on the direct collapse of over-dense regions as they enter the horizon. 
For a recent review of PBH formation and evolution see, e.g. \cite{Shankaranarayanan:2026hnn}.}

In this scenario, PBHs should only form when an over-density re-enters the cosmological horizon with compaction larger than some threshold value $\delta_\mathcal{C}$. 
This threshold was first calculated by Carr \cite{Carr:1975qj}, who used a Newtonian description to argue that over-densities must be larger than the Jeans length but smaller than the particle horizon. This results in a peak density contrast (at scales smaller than the horizon) at least equal to the equation of state parameter $w = p/\rho$ of the perfect fluid that fills the Friedmann-Lema\^itre-Robinson-Walker (FLRW) universe. 
For a radiation fluid, this is $\delta_c \simeq 1/3$. 
Later, Harada, Yoo and Kohri \cite{Harada:2013epa} estimated the threshold using general relativity instead, finding $\delta_c \sim 0.4$ for radiation domination. 
Numerical and analytical techniques have also found that the threshold, $\delta_c$ depends on the shape of the perturbation \cite{Polnarev:2006aa, Musco:2012au, Musco:2018rwt, Escriva:2019phb, Musco:2020jjb, Ianniccari:2024bkh}. 
For an initial profile that is characterized by a single parameter, the authors of \cite{Musco:2018rwt, Escriva:2019phb} estimate a bound of $0.4 \leq \delta_c \leq 2/3$ depending on how sharply peaked the profile is. 

One open question is whether there is a universal threshold. 
The authors of \cite{Escriva:2019phb} argue that the threshold of an initially spherically symmetric fluctuation given in terms of the compaction function averaged over the inside of a sphere of radius $r_m$ (the radius at which the compaction function is maximized) is approximately a universal threshold. 
They further posit that the threshold of the volume averaged compaction function depends only on the curvature at the maximum of the compaction.
For a radiation-dominated universe this threshold is $\delta_c \sim 0.4$ \cite{Escriva:2019phb}, and an analytic argument for this result is provided in \cite{Kehagias:2024kgk}. 
For a more detailed review of the threshold criteria and PBH formation via direct collapse see \cite{Escriva:2021aeh}. 

In \cite{Musco:2020jjb}, the authors calculate the threshold during radiation domination; the authors take into account both nonlinearities between the curvature perturbation and the density contrast, as well as nonlinear effects arising at cosmological horizon crossing. 
They found that if the threshold is calculated around the time when PBHs form (i.e.~at the time of cosmological horizon crossing), then the range of thresholds across all possible profile shapes is $0.7 \leq \delta_c \leq 1.15$ \cite{Musco:2020jjb}.  

In this paper, we present results from numerical relativity simulations that examine the collapse of super-horizon over-densities in a relativistic radiation fluid in the early Universe.  
The simulations here extend previous works in several key ways.  First, we employ periodic, toroidal boundary conditions; we do not assume a cosmology at the boundary of the simulation, instead allowing the matter content to drive the dynamics of the Universe self-consistently.\footnote{
Several previous numerical works that simulate scalar fields~\cite{Yoo:2018pda,Yoo:2018dni}, black hole spacetimes~\cite{Yoo:2013yea}, or radiation fluids~\cite{Yoo:2020lmg} 
on a lattice also employ boundary conditions that mimic periodic space, they evolve an octant of the full periodic domain and impose reflective boundary conditions in all directions \cite{Yoo:2013yea,Yoo:2018pda,Yoo:2018dni,Yoo:2020lmg}.  
These works also set the outer region of the grid to be homogeneous and isotropic, which plays the main role in driving the large-scale expansion.
Our set up is less constrained, but we find that in practice our results are nevertheless consistent with the previous results.}
Whilst our initial over-density is spherically symmetric, the Cartesian nature of our grid and toroidal topology break that symmetry, allowing 
us to test whether relaxing the exact symmetry changes the result.
We utilize the BSSN formalism (in contrast to the Misner-Sharp \cite{Misner:1964je} or Hernandez-Misner \cite{Hernandez:1966zia} formalisms that have been employed in the majority of PBH formation simulations), which allows departures from spherical symmetry, has the potential to provide a more stable long-term evolution beyond horizon formation, and allows us to check the slicing dependence of the threshold for collapse. 
In future work we will use this to study the subsequent black hole evolution and accretion. 

We also present a scheme for choosing constraint-satisfying initial data that do not rely on the gradient expansion \cite{Shibata:1999zs,Harada:2015yda,Salopek:1990jq}. We balance the intrinsic curvature of the initial slice with the extrinsic curvature corresponding to the asymptotic expansion of the Universe and the over-density, following the treatment in \cite{Baumgarte:2025vvs}, to identify limits on the initial data and construct both weak- and strong-field branch solutions. However, we restrict our initial data to the {\sl type-I class} of over-densities \cite{Yoo:2018kvb}, i.e.~to data for which the areal radius remains a monotonically increasing function of the radial coordinate. Consequently, we are able to study initial data from the strong-field branch without the ambiguities that can arise in {\sl type-II class} scenarios that are already highly likely to collapse from geometric arguments \cite{Kopp:2010sh}.

This paper is organized as follows. 
In \Sec{sec:setup} we describe the numerical set up, including a detailed discussion of the initial conditions, numerical formalism, and the treatment of the fluid model. 
We describe our numerical simulations in \Sec{sec:numerics}, including slicing conditions and diagnostics for PBH formation.  We present results in \Sec{sec:results}.  We work in natural units where $c=\hbar=1$, however, we retain a dimensionful Newton's constant, $G = \mpl^{-2}$.

\section{The Setup} 
\label{sec:setup}

\subsection{Einstein's Equations} 
\label{subsec:EEs}

We employ the Baumgarte-Shapiro-Shibata-Nakamura (BSSN) scheme to conduct simulations of fully nonlinear general relativity on a finite domain using periodic boundary conditions.
In the BSSN formalism, spacetime is foliated into a family of 3-dimensional spacelike hypersurfaces given by slices of constant scalar time function $t$, which serves as a coordinate time labeling the slices.  The metric is generally written in the Arnowitt-Deser-Misner (ADM) form \cite{Arnowitt:1959ah}
\begin{equation}\label{eq:BSSNmetgeneral}
        g_{\mu \nu}= \left[\begin{array}{cc} -\alpha^2 + \beta_k\beta^k & \beta_j \\ \beta_i & \gamma_{ij}\end{array}\right], 
\end{equation}
where $\alpha$ is the lapse function and $\beta_i$ is the shift vector. 
In this 3+1 decomposition, the spatial surfaces of the simulation are defined by a time-like normal vector $n_\alpha = (-\alpha, 0,0,0)$.

We follow the convention that the spatial metric $\gamma_{ij}$ is rewritten in terms of a conformal factor $\psi = e^\phi$ and a unit-determinant spatial metric $\bar{\gamma}_{ij}$, such that $\gamma_{ij} = \psi^4\bar{\gamma}_{ij}= e^{4\phi}\bar{\gamma}_{ij}$.  Likewise, we decompose the extrinsic curvature, $K_{ij}$, into its trace, $K$, which we refer to as the mean curvature, and a trace-free $\tilde{A}_{ij}$, such that $K_{ij} = e^{4\phi}\tilde{A}_{ij} + \gamma_{ij}K/3$.  The 3+1 decomposition of Einstein's equations produces evolution equations for these quantities from one spatial hypersurface to the next, as well as constraints on their values within a single spatial slice. 
In an extension to the ADM formalism, BSSN introduces conformal connection functions as dynamical variables, $\bar{\Gamma}^i \equiv \bar{\gamma}^{jk} \bar{\Gamma}^{i}_{jk}$. 
These variables evolve independently, creating a well-posed system for numerical evolution. (Note that the resulting constraint $\bar{\Gamma}^i + \partial_j \bar{\gamma}^{ij} = 0$ is not explicitly imposed during the evolution).
In addition to the evolution equations for $\psi, \bar{\gamma}_{ij}, \tilde{A}_{ij}, K,$ and $ \bar{\Gamma}^i$, the BSSN formalism includes two constraint equations which must be satisfied throughout the simulation to produce valid solutions to Einstein's equations. 
See \cite{Baumgarte:2010ndz, Baumgarte:2021skc} for a more detailed review of the BSSN formalism. 

\subsection{Relativistic Hydrodynamics} 
\label{subsec:model}

We consider a universe in which the only component is a perfect fluid; see \App{sec:appendixB} for more details on the derivation of the fluid variables and a discussion of program units.
The stress energy tensor for this fluid is
\begin{equation}
\label{eq:stressfull}
        T^{\mu \nu} = \left(\rho_0 + \frac{p}{c^2} \right) U^{\mu} U^{\nu} + p g^{\mu \nu},
\end{equation}
where $\rho_0$ is the rest-frame energy density, $p$ is the isotropic pressure, and $U^\mu$ is the four velocity of the fluid. We define the Lorentz factor $W \equiv - n_\mu U^{\mu} =\alpha U^t$ of the fluid relative to the normal vector, $n_\alpha$, as well as the three-velocity 
\begin{equation}
v^i \equiv \frac{1}{W}{\gamma^{i}}_\alpha U^\alpha = \frac{U^i}{W} + \frac{\beta^i}{\alpha}.
\end{equation}
Since we are interested in the formation of PBHs from inflationary fluctuations that re-enter during the radiation-dominated era, we assume a radiation fluid with equation of state parameter $w \equiv p/\rho_0 = 1/3$, for which the stress-energy, \Eq{eq:stressfull}, reduces to
\begin{equation}
        T^{\mu \nu} = \frac{4}{3} \rho_0 U^{\mu} U^{\nu} + \frac{\rho_0}{3} g^{\mu \nu}. 
\end{equation}

We write the relativistic equations of motion for the fluid, $\nabla_\alpha T^{\alpha \beta} = 0$, in terms of the conformally-related internal energy density $E \equiv \sqrt{\gamma} \, n_\alpha n_\beta T^{\alpha \beta} $ and momentum density $P^i \equiv - \sqrt{\gamma} \, n_\nu T^{i \nu}$ to obtain
\begin{align}
\label{eqn:Eom_E_2}
\partial_t E 
&+ \alpha\, (\partial_i P^i)
 - \beta^i (\partial_i E)
 - E\, (\partial_i \beta^i)  \nonumber\\
&= -2 P^i (\partial_i \alpha)
   + P^i \beta^j K_{ij}
   + K_{ij} S^{ij}_{\rm v}
   - P^j \beta^i K_{ij},
\end{align}
and 
\begin{align}
\label{eqn:Eom_P_2}
\partial_t P_j 
&+ \gamma_{kj}\, \partial_i S^{ik}_{\rm v}
 + S^{ik}_{\rm v} (\partial_i \gamma_{kj})
 - P_j (\partial_i \beta^i)
 - \beta^i (\partial_i P_j)  \nonumber\\
&= - E\, (\partial_j \alpha)
   + \frac{P^i}{2}\, \beta^k (\partial_j \gamma_{ik})
   + P_k (\partial_j \beta^k)  \nonumber\\
&\quad
   + \frac{1}{2} S^{ik}_{\rm v} (\partial_j \gamma_{ik})
   - \frac{1}{2} P^k \beta^i (\partial_j \gamma_{ik}).
\end{align}
where the conformally-related stress, $S^{ij}_{\rm v}$, is defined by \Eq{eq:sijcode} (see \App{sec:appendixB} and \App{sec:appendixC} for more details).

\subsection{Initial Conditions}
\label{sect:initialconditions}
One of the great challenges of performing numerical simulations in full numerical relativity is defining physically well-motivated initial data. 

In linear cosmological perturbation theory, physical perturbations can be characterized by gauge-invariant combinations of the metric and matter perturbations, see e.g.~\cite{Mukhanov:1990me,Baumann:2009ds} for pedagogical reviews. In the standard PBH formation mechanism, quantum fluctuations of these gauge-invariant quantities are {\sl frozen out} during inflation at the time when each mode exits the cosmological horizon \cite{Hodges:1990bf}.  In many models, then, the superposition of these modes leads to large, local curvature fluctuations.  Here, following the treatment in \cite{Wands:2000dp,Lyth:2004gb}, we define a nonlinear curvature fluctuation $\zeta$ to realize an initial over-dense region and write the metric surrounding this over-dense region in spherical polar coordinates centered at the maximum value of $\zeta$ as
\begin{equation}
    ds^2 = -dt^2 + a^2e^{2\zeta}\left[dr^2 + r^2 d\Omega^2\right].
    \label{eq:metricnonpertwzeta}
\end{equation}
Here 
\begin{equation}
\label{eq:defofzeta}
    \zeta = Ae^{-r^2/2\sigma^2}
\end{equation} 
is assumed to be a spherical Gaussian, in terms of the coordinate radius, $r$, as expected by studying peak theory \cite{Bardeen:1985tr, Musco:2020jjb}.\footnote{In linear perturbation theory, $\zeta$ corresponds to the standard gauge-invariant quantity called the {\sl comoving curvature perturbation on constant-density hypersurfaces} and can be expressed as
\begin{equation}
    \zeta = -\Psi - \frac{\bar{\rho}}{\dot{\bar{\rho}}}\delta \rho,
\end{equation}
where $\bar{\rho}$ is the energy density of the corresponding {\sl background} spacetime. 
In the $\delta\rho\rightarrow0$ perturbation theory limit, the metric becomes
\begin{equation}
    ds^2 = -dt^2 + a^2\left(1+2\zeta\right)\left[dx^2+dy^2+dz^2\right].
    \label{eq:metricpertwzeta}
\end{equation}
Since PBHs are formed when there are large, statistically rare excursions of the $\zeta$ field, $\zeta$ is allowed to be a nonlinear quantity \cite{Shibata:1999zs}.
}

In spherical symmetry, the {\sl areal radius}, $R = r\psi^2$, provides a local measure of the size of the space at a coordinate radius $r$, by relating it to the proper area of the coordinate sphere. It can be used to define the {\sl background mass}, $\bar{M}$, which is the amount of mass that would be enclosed in an areal radius $R$ in the corresponding FLRW spacetime with the same (but homogeneous) asymptotic density $\bar{\rho}$, 
\begin{equation}
    \bar{M} = \frac{4}{3}\pi \bar{\rho} R^3. \label{eq:Mbar}
\end{equation}
Note that in inhomogeneous spacetimes, the areal radius can be different from the proper radius (i.e. the one obtained by integrating the proper distance on the spatial slice from $r=0$ to $r$), and in such cases the areal radius may not accurately represent the background mass; however, in all the cases we consider here, the areal radius increases monotonically with the proper radius and differs by an $\mathcal{O}(1)$ factor, and so it remains a meaningful quantity.  We also note that in a $K=0$ slicing, a turnaround in areal radius would guarantee subsequent BH formation, as explained in \cite{Kopp:2010sh}.

Choosing the scale factor $a_*=1$ on our initial slice together with $\bar{\gamma}_{ij} = \delta_{ij}$, we can identify the initial value of the conformal factor as
\begin{equation}
    \psi_* = e^{\zeta_*/2}.
\end{equation}
Throughout this work, we use an asterisk subscript to define quantities on the initial surface of the simulation, e.g.~$\psi_*$.   

The choice of initial $\zeta_*$, which in turn determines the spatial metric, must be consistent with the {\sl Hamiltonian constraint} \cite{Arnowitt:1959ah,Mukhanov:1990me,Baumann:2009ds},
\begin{equation}
    \bar \nabla^2 \psi_* - \frac{1}{12} \psi_*^5 K_*^2 + \frac{1}{8} \psi_*^{-7} \tilde A_{ij\,*} \tilde A^{ij}_* = - \frac{2 \pi}{m_{\rm pl}^2} \psi_*^5 \rho_*.
    \label{eq:hamful}
\end{equation}
The first term in Eq.~\eqref{eq:hamful} is set by our choice of $\zeta_*$, but we can make several choices for the other terms in the constraint.  
In a homogeneous and isotropic spacetime the trace of the extrinsic curvature is related to the Hubble parameter, $H=\dot{a}/a$, by $K= -3H$.  To recover this relation asymptotically, we set the constant value
\begin{equation}
    K_* = -\sqrt{\frac{24\pi \bar{\rho}_*}{m_{\rm pl}^2}}.
    \label{eq:Kscalar}
\end{equation}  
We choose the fluid to be initially at rest so that the fluid momentum density vanishes, $\mathcal{P}^i =0$. Since the gradients of $K_*$ vanish, the momentum constraint is trivially satisfied with the choice  $\tilde{A}_{ij*} = 0$. 

\begin{figure}[t]
    \centering
    \includegraphics[height=5in]
    {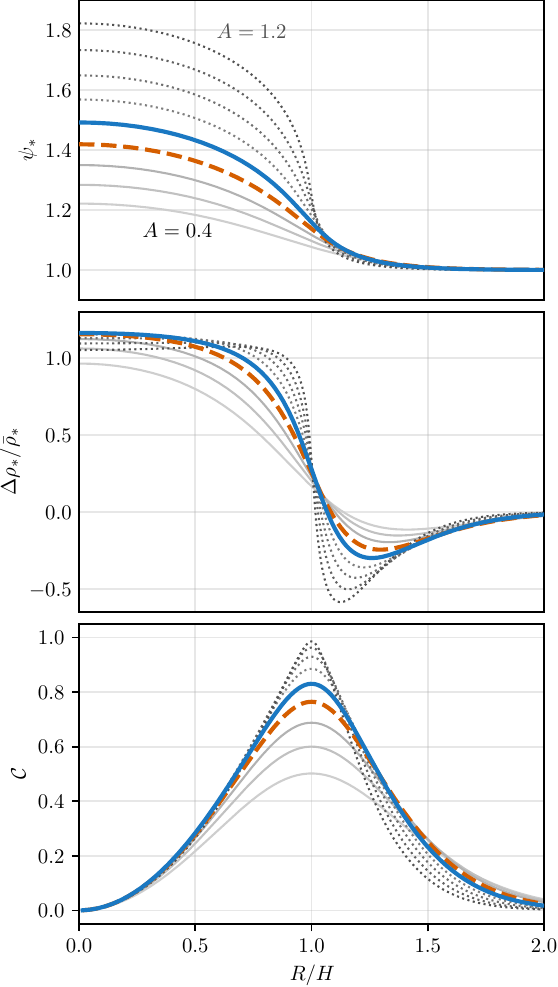}
    \caption{The initial conformal factor, $\psi_*$ (top panel), the initial density contrast, $\Delta \rho_*/\bar{\rho}_*$ (middle panel) and the compaction, $\mathcal{C}$ (bottom panel) as a function of the areal radius, $R$, as described in \Sec{sect:initialconditions} and  Table~\ref{tab:initialdataparams}.  The thicker solid (blue) and dashed (red) lines identify cases A and B, the two border cases discussed in \Sec{sec:results} (see Table~\ref{tab:initialdataparams}). Notably, these bracket the maximum central value of $\Delta \rho_*/\bar{\rho}_*$, beyond which increasing the amplitude of the perturbation in the conformal factor decreases the corresponding over-density, as seen in \Fig{fig:Avsdelta}.}
    \label{fig:ICpsirhoMSMass}
\end{figure}

Our choices for the curvature terms fully determine $\rho_*$ from \Eq{eq:hamful}. We parameterize this as a deviation from the asymptotic value, $\Delta \rho_* = \rho_*-\bar{\rho}_*$, for which \Eq{eq:hamful} yields
\begin{equation}
    \frac{\Delta \rho_*}{\bar{\rho}_*} = -\frac{4}{3H_*^2}\frac{\nabla^2\psi_*}{\psi_*^5}.
\end{equation}
Note that $\Delta \rho_*$ does not need to be small. However, one can show that there is a limit on the size of $\Delta \rho_*/\bar{\rho}_*$, despite there being no limit on the amplitude $A$ of $\zeta$ in \Eq{eq:defofzeta}, because of the competition between the increasing Laplacian in the numerator and the $\psi^5$ term in the denominator \cite{Baumgarte:2006ug,Baumgarte:2025vvs}.  \Fig{fig:ICpsirhoMSMass} shows the size of $\Delta \rho_*$ for the range of parameters studied below.

We use the Misner-Sharp mass \cite{Misner:1964je,Misner:1973prb,Thornburg:1998cx} as a measure of the total mass enclosed inside a given coordinate radius, $r$,
\begin{align} \label{MS_mass}
    \frac{M_{\rm MS}(r)}{m_{\rm pl}^2} &=  \frac{R}{2}\left(1-\left(\nabla^\mu R \right)\left(\nabla_\mu R\right)\right) \nonumber\\
    &=\frac{r^3 \psi^6 K^2}{18} - 2r^3(\partial_r \psi)^2 - 2r^2 \psi \partial_r \psi.
\end{align}
This allows us to define the {\sl compaction} \cite{Musco:2020jjb} as 
\begin{equation} \label{eq:compact}
    \mathcal{C} \equiv \frac{2G\delta M}{R} = \frac{2}{m_{\rm pl}^2}\frac{M_{\rm MS}-\bar{M}(R)}{R},
\end{equation}
where $\bar M$ is the background mass as defined in \Eq{eq:Mbar}.
This provides a measure of whether the over-density is large enough to satisfy the hoop conjecture \cite{Misner:1973prb}, which occurs when $\mathcal{C} \geq 1$, although an evolution of these initial data is nevertheless required to see if a black hole results.

Following~\cite{Musco:2020jjb}, we use the compaction function to define the characteristic scale of the over-density. For given values of the amplitude $A$ and width $\sigma$, the compaction $\mathcal{C}(R)$ exhibits a maximum value at some areal radius, $R_c$, which we identify with the size of the over-dense region.  

We initialize our simulations at the time of cosmological horizon crossing, defined by
\begin{equation}
    R_c H_* = 1.
\end{equation}
Rather than expressing the threshold in terms of the density contrast, we employ the commonly-used definition $\delta_c = \mathcal{C}(R_c)$ evaluated at the initial time.

\begin{figure}[t]
    \centering
    \includegraphics[width=\columnwidth]{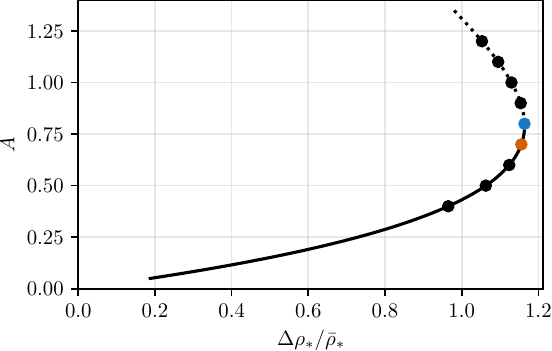}
    \caption{A plot of $A$ versus  $\Delta\rho_*/\bar\rho_*$ (evaluated at $r=0$) for the initial data described in \Sec{sect:initialconditions}. We see that increasing $A$ only increases the over-density up to a maximum value.  The dots represent the specific simulations presented in this work; the red and blue dots identify the two border cases discussed in \Sec{sec:results}.}
    \label{fig:Avsdelta}
    \end{figure}

To construct initial data, we consider a set of values of $A$, corresponding to increasingly large excursions of $\zeta_*$.
We choose a width $\sigma$ for every value of $A$ so that the maximum of the compaction function satisfies $R_c H_*=1$ on the initial surface.  \Fig{fig:ICpsirhoMSMass} shows the radial profile of $\psi_*$, $\Delta \rho_*/\bar{\rho_*}$, and $\mathcal{C}$ for this family of solutions. We can also look at these initial data in terms of the two branches discussed in \cite{Baumgarte:2006ug,Baumgarte:2025vvs}, see \Fig{fig:Avsdelta}. In this figure we see that increasing $A$ only increases the over-density up to a maximum value, and that the cases studied lie around that maximum, with cases on both the strong-field (upper) and weak-field (lower) branch. Whilst cases on the lower, weak-field branch may have the same value of $\Delta \rho_*/\bar{\rho}_*$ at $r=0$ as corresponding cases on the upper, strong-field branch, they are physically distinct, as can be seen in \Fig{fig:ICpsirhoMSMass}. Therefore, the black hole formation threshold we identify is not a threshold in $\Delta \rho_*/\bar{\rho}_*$, but rather a threshold in $A$, the curvature excursion. In order to support initial data with higher central $\Delta \rho_*/\bar{\rho}_*$, one would necessarily need to relax the assumption of an initial constant mean curvature $K_*$.

\section{Numerical Methods} 
\label{sec:numerics}
 
\subsection{GABERel}
\label{sec:GABERel}
We use GABERel \cite{Giblin:2019nuv,Adshead:2023mvt}, a version of GABE \cite{Child:2013ria} that solves the fully nonlinear Einstein's equations alongside the equations for relativistic hydrodynamics as discussed in \Sec{sec:setup}. 

GABERel creates a cubic grid of $N\times N\times N$ gridpoints, at which the dynamical degrees of freedom are stored. We impose periodic boundary conditions in each Cartesian direction.  We place the center of the over-dense region at the corner of the box -- and hence all eight corners -- because it makes calculating distances from the center of the over-dense region more convenient.  The point in the middle of the box, located at $x=y=z=L/2$ is, therefore, the furthest point from the center of the over-dense region in any direction; this point will be used as a reference point at which we calculate (approximately) asymptotic reference values, to which we will refer with subscripts $\infty$.  This is justified since the center point of the box remains locally homogeneous, with spatial gradients of the dynamical variables remaining small throughout the simulations.

In the simulations presented here, we employ a lattice with $N = 256$ points and comoving size $ L = 8/H_*$ or $L = 9/H_*$ for simulations with values of $\sigma$ that range from $\sigma \approx  0.45/H_*$ to $\sigma \approx  0.61/ H_* $.  The full set of parameters we use are listed in Table~\ref{tab:initialdataparams}. 
\begin{table}[]
    \centering
    \renewcommand{\arraystretch}{1.6}
    \setlength{\tabcolsep}{12pt}
    \begin{tabular}{|c|c|c|c|l}
    \cline{1-4}
         $\delta_c$ & $A$ & $\sigma$ & $L$ & \\
         \cline{1-4}
         \cline{1-4}
         .50 & .4 & $.61/H_*$ & $9/H_*$ & \\
         .60 & .5 & $.68/H_*$ & $9/H_*$ & \\
         .69 & .6 & $.56/H_*$ & $9/H_*$ & \\
         .76 & .7 & $.54/H_*$ & $9/H_*$ & Case B \\
         .83 & .8 & $.52/H_*$ & $8/H_*$ & Case A \\
         .89 & .9 & $.50/H_*$ & $8/H_*$ & \\
         .93 & 1.0 & $.48/H_*$ & $8/H_*$ & \\
         .96 & 1.1 & $.47/H_*$ & $8/H_*$ & \\
         .99 & 1.2 & $.45/H_*$ & $8/H_*$ & \\
    \cline{1-4}
    \end{tabular}
    \caption{Parameters that determine the initial conditions for our family of simulations.}
    \label{tab:initialdataparams}
\end{table}

We choose box sizes that are sufficiently large so that the gradients of all fields are below machine precision halfway across the box, yet small enough to sufficiently resolve the scales of the initial over-densities. The spherical symmetry of the overdensity is broken by the periodic boundaries, but we do not see any deviation from homogeneity at points in the box that are more than a few $H_*^{-1}$ from the over-dense region.  
While we are not {\sl enforcing} an asymptotic region of homogeneity and isotropy to drive the evolution, we find no inconsistency with previous work that does. 

We use a standard fourth-order Runge Kutta method for time integration, and employ a timestep set by both the resolution and length of the box, as $\Delta t = L/(80N)$. 
We also performed a set of tests to confirm that the physical conclusions of this work are independent of our choices of numerical parameters: we ran our family of simulations for two different spatial resolutions, $N = 128$ and $N=256$, as well as a set of increasingly larger time steps. 
As expected, constraint violation improved for both higher resolutions simulations and those with smaller time steps. 
\begin{figure*}[ht]
    \centering
    \includegraphics[width=\textwidth]
    {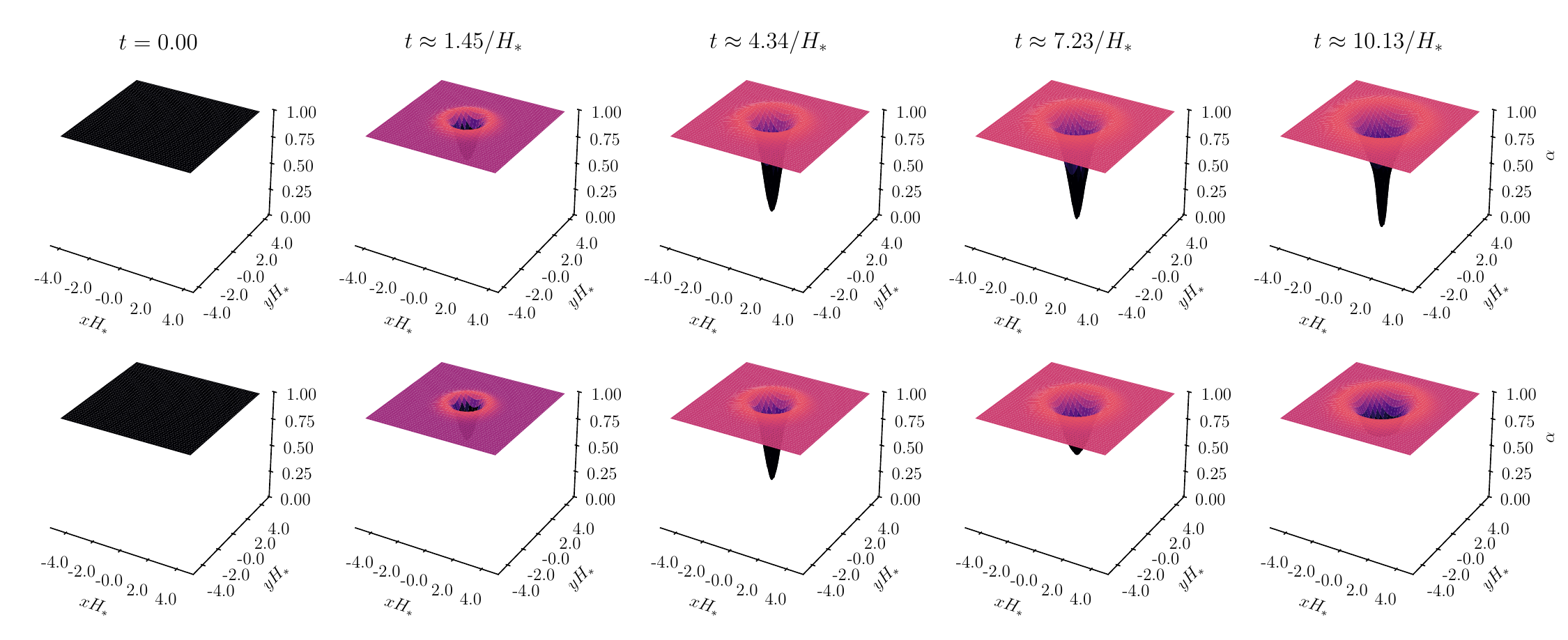}
    \caption{Two-dimensional slices of the lapse, $\alpha$, at $t=0$, $t\approx1.45/ H_*$, $t\approx 4.34/H_*$, $t\approx7.23/H_*$, $t\approx 10.13/ H_*$ (from left to right) for Case A (top row) and Case B (bottom row). The simulation for case A, which exceeds the compaction threshold for formation, has a steadily decreasing lapse at $r=0$ which approaches zero at the end of the simulation. The simulation corresponding to case B, which is below the identified critical threshold, shows an initially decreasing lapse value which then increases back towards one by the end of the simulation. }
    \label{fig:alpha2d}
\end{figure*}

\subsection{Slicing Conditions}

In the BSSN formalism, the lapse function, $\alpha$, and the shift vector, $\beta^i$, encode gauge freedom, and may be freely specified. 
It is common to adopt the ``Bona-Mass\'o'' slicing condition in numerical relativity 
    \begin{equation}
        (\partial_t - \beta^i \partial_i) \alpha = - \alpha^2 f(\alpha) K,
    \end{equation}
where $f(\alpha)$ is a specific Bona-Mass\'o function chosen for a particular scenario \cite{Bona:1994dr}.
For instance, the choice of $f(\alpha) = 1$ results in harmonic slicing \cite{Gourgoulhon:2007ue}.
Another common choice is  $1 + \log$ slicing condition, $f(\alpha) = 2/\alpha$.
We found reliable, stable evolution in choosing  
    \begin{equation}
        \dot{\alpha} = - \frac{1}{3} \alpha \left( K - K_\infty \right) + \beta^i \partial_i \alpha 
        \label{eq:lapseslice}
    \end{equation}
where $K_\infty$ is the mean curvature evaluated at the center of the box.
We found that the choice $\beta = \dot{\beta} = 0$ represents a compromise between stable evolution for cosmological spacetimes, while still allowing the simulations to form apparent horizons. This choice means that the spatial coordinates of the normal observer do not change.\footnote{While the coordinates of the normal observer do not change, the non-constant lapse will result in accelerated normal observers, so normal observers are not geodesic observers in this gauge choice.  We tested a variety of slicing conditions, including the typical ``moving puncture'' condition in black holes simulations, which use a ${1+\log}$ slicing condition for $\alpha$ and a hyperbolic Gamma driver condition for the shift, $\beta$. We also tried shock avoiding slicing conditions, like the one described in \cite{Alcubierre:1996su, Alcubierre:2002iq,Baumgarte:2022ecu}, but we found the most stable evolution resulted from the slicing conditions given above.}

For some of the simulations, coordinate shocks develop after the formation of the PBH; our slicing conditions, \Eq{eq:lapseslice} and $\beta=0$, cannot resolve these for long. There are known fixes to resolve these shocks and permit the simulations to last longer; here, we are interested in black hole formation and so simply allow the simulation to end when shocks appear. 

\subsection{Condition for Primordial Black Hole Formation}

\begin{figure*}[t]
    \centering
    \includegraphics[height=3 in]
    {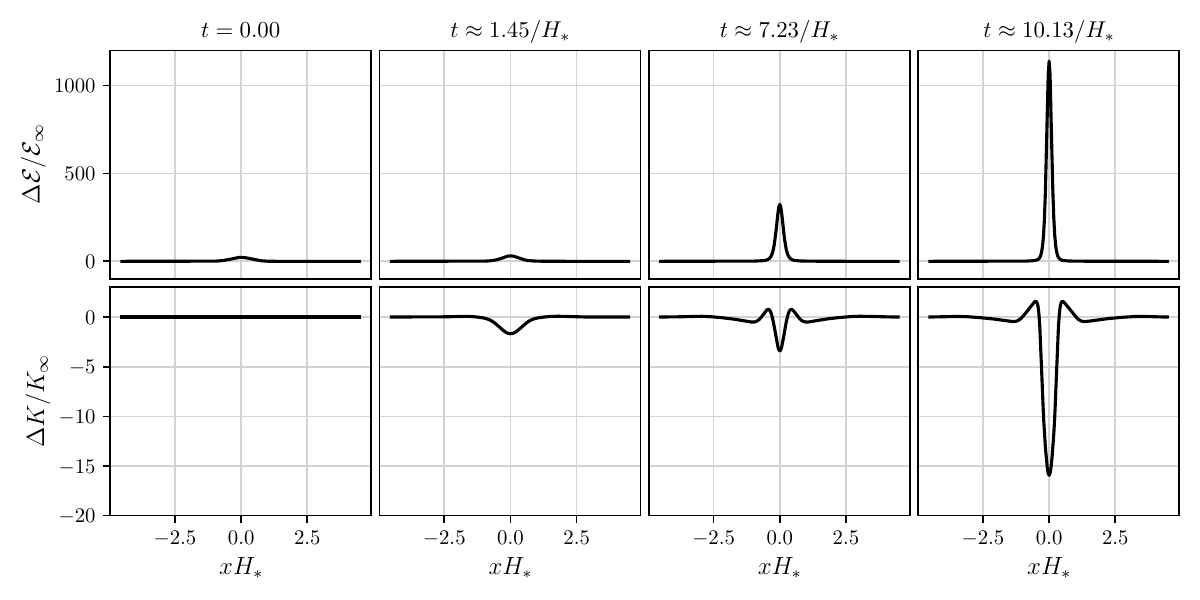}
    \caption{One-dimensional slices of the energy density contrast (top row) and mean curvature contrast (bottom row) at $t=0$, $t\approx1.45/ H_*$, $t\approx7.23/H_*$, and $t\approx 10.13/ H_*$ (left to right) for case A, which has an initial compaction at horizon crossing of $\mathcal{C}=0.83$. The energy density contrast increases over time at $r=0$, while the mean curvature contrast drops below negative one, indicating that the mean curvature itself is negative and a region of space is contracting as opposed to expanding. This is consistent with collapse and black hole formation. We calculate the asymptotic values $E_\infty$ and $K_\infty$ at the center of the box as described in \Sec{sec:GABERel}.}
    \label{fig:a0.9Ephi}
    \end{figure*}

The threshold for PBH formation has been calculated as $0.7\leq \delta_c \leq 1.15$ in \cite{Musco:2020jjb} for a variety of spherically-symmetric profiles when embedded in a radiation-dominated background.
Over-densities whose compaction exceeds $\delta_c$ at the time of cosmological horizon crossing {\sl should} form a black hole.
To find this threshold, we need to determine when, and if, black holes form in our simulations.
There are several available diagnostics for black hole formation, including the formation of an apparent horizon, a trapped surface on a spatial slice, the collapse of the lapse function, or the tracking of an event horizon via post-processing integration of null geodesics, among others \cite{Thornburg:2003sf,Baumgarte:2010ndz}.
Here we rely on the identification of an apparent horizon to determine whether a black hole forms; however, we also demonstrate how this is coincident with a local decrease in the lapse for our choice of slicing condition.

Unlike event horizons, which rely on global properties of spacetime, apparent horizons can be identified within a spatial slice. 
The presence of an apparent horizon guarantees the formation of an event horizon and its area provides a lower bound for the resulting black hole mass.
To search for apparent horizons, we follow the  procedure outlined by Thornburg \cite{Thornburg:2003sf}. 
The apparent horizon is the outermost marginally trapped surface, i.e.~a spacelike 2-D surface, $\mathcal{S}$, for which the expansion, $\Theta$, of outgoing normal null geodesics vanishes.  Denoting $N^i$ as the (spatial) unit-length outward pointing normal vectors to $\mathcal{S}$, the expansion is given by
    \begin{equation}
        \Theta = \nabla_i N^i + K_{ij} N^i N^j - K = 0. 
    \end{equation}
For spherical surfaces, centered at the origin, the $N^i$ can be written as
    \begin{equation}
        N^i = \frac{g^{ij}s_j}{(g^{kl} s_k s_l )^{1/2}},
    \end{equation}
where the
\begin{equation}
    s^i = \frac{x^i}{\sqrt{x^2+y^2+z^2}}
\end{equation}
are the non-unit normal vectors to $\mathcal{S}$.

In terms of 3+1 variables, we write the expansion $\Theta$ as 
    \begin{equation} 
        \Theta = \frac{A}{D^{3/2}} + \frac{B}{D^{1/2}} + \frac{C}{D} - K,
        \label{eq:expansion}
    \end{equation}
where we have used the abbreviations
    \begin{subequations}
        \begin{align}
        A &= - (g^{ik} s_{k}) (g^{jl} s_{l}) \partial_i s_j - \frac{1}{2} (g^{ij} s_j) \left[(\partial_i g^{kl}) s_k s_l \right] \\
        B &= (\partial_i g^{ij}) s_j + g^{ij} \partial_i s_j + (\partial_i \ln \sqrt{g}) (g^{ij} s_j)\\
        C &= K^{ij} s_i s_j\\
        D &= g^{ij} s_i s_j. 
    \end{align}
    \end{subequations}
    
To search for apparent horizons, we define a family of 2-spheres centered on the corner(s) of the grid, i.e.~the center of the over-dense region, and evaluate the expansion as a function of the coordinate radius of these spheres (for an illustration, see \Fig{fig:ah_example} below).

\section{Results} 
\label{sec:results}

\begin{figure*}[t]
    \centering
    \includegraphics[height=3 in]
    {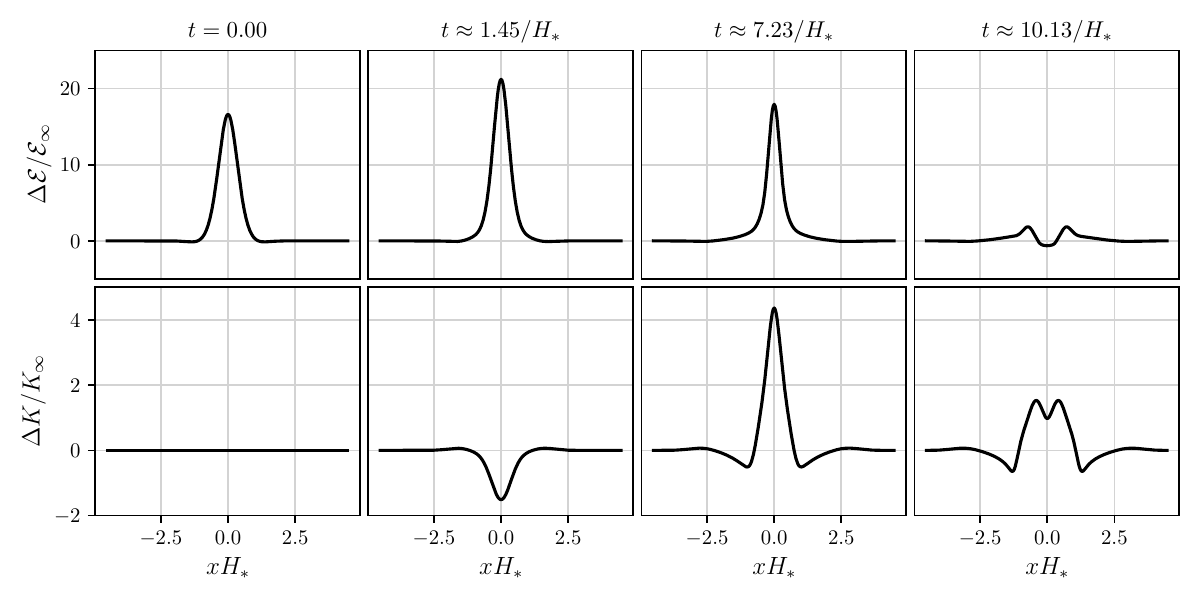}
    \caption{One-dimensional slices of the energy density contrast (top row) and mean curvature contrast (bottom row) at $t=0$, $t\approx1.45/ H_*$, $t\approx7.23/H_*$, and $t\approx 10.13/ H_*$ (left to right) for case B, which has an initial compaction at cosmological horizon crossing of $\mathcal{C}\approx0.77$. While the peak energy density contrast initially increases at the center of the initial over-dense region, it then proceeds to decrease back towards zero. Early in the simulation, the extrinsic curvature contrast dips at the center of the over-dense region before returning to zero over the course of the simulation. We calculate the asymptotic values $E_\infty$ and $K_\infty$ at the center of the box as described in \Sec{sec:GABERel}.}
    \label{fig:a0.7Ephi}
\end{figure*}

Having run the simulations of spherical, horizon-sized over-densities with the set up described above, we confirm PBH formation in collapsing cases by identifying apparent horizons. We determine the threshold value for the compaction at cosmological horizon crossing to be $0.77<\mathcal{C}<0.83$.  
We find that in simulations with initial compactness lower than our threshold range, the over-dense region in the fluid energy density dissipates. In contrast, in simulations above this threshold range, the fluid energy density grows at the center of the over-density, the lapse function vanishes at the center of the over-dense region, and the expansion shows the existence of an apparent horizon.

We present more detailed results from two simulations that bracket the threshold for PBH formation: one in which the over-dense region collapses and forms an apparent horizon, which we refer to as case A (for {\sl a}bove the threshold) and one in which the over-dense region dissipates which we label as case B (for {\sl b}elow the threshold). Case B, in which the over-dense region dissipates, has an initial compactness value at cosmological horizon crossing of 
$\mathcal{C}=0.77$,
while case A -- the simulation which forms an apparent horizon -- has an initial compactness of $\mathcal{C}=0.83$. \Fig{fig:alpha2d} shows the evolution of the lapse in these two cases. For case A, the lapse steadily decreases, approaching zero at the center of the over-density. By contrast, the lapse in case B decreases for a short time before returning to unity for the remainder of the simulation.

When using Bona-Mass\'o slicing conditions, the lapse may act as an indicator of the formation of compact objects, another indicator is the growth of the density. \Fig{fig:a0.9Ephi} and \Fig{fig:a0.7Ephi} show the behavior of the fluid energy density contrast,
\begin{equation}
    \frac{\Delta \mathcal{E}}{\mathcal{E}_\infty} \equiv \frac{\mathcal{E}-\mathcal{E}_\infty}{\mathcal{E}_\infty},
\end{equation}
and mean curvature contrast,
\begin{equation}
    \frac{\Delta K}{K_\infty} \equiv \frac{K-K_\infty}{K_\infty},
\end{equation}
for both cases A and B.
For case A, the over-dense region begins to collapse and eventually forms an apparent horizon. The energy density contrast in this case steadily increases over the course of the simulation. As the region collapses, we also see that the mean curvature contrast is negative near the center of the over-density, indicating that the comoving volume is decreasing. In case B, the energy density contrast initially grows, but then decreases as it becomes more homogeneous; the mean curvature contrast also approaches homogeneity toward the end of the simulation.

\begin{figure}[h!]
    \centering
    \includegraphics[width=\columnwidth]
    {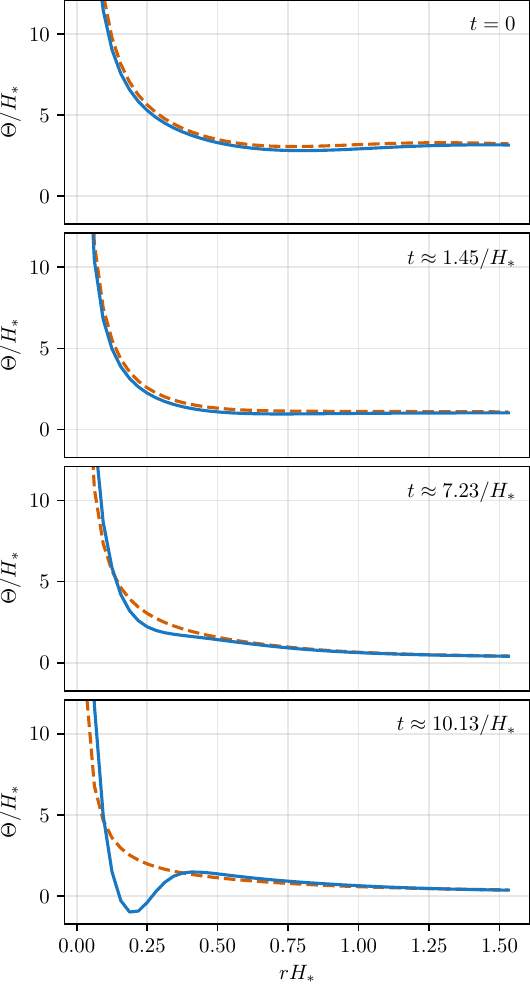}
    \caption{The expansion, $\Theta(r)$ as calculated on a set of spacelike 2-spheres of varying coordinate radius, $r$, centered on the over-dense region for the two reference cases, Case A, $\mathcal{C} \approx 0.83$ (blue), and Case B,
    $\mathcal{C} \approx 0.77$ (red, dashed),
    evaluated (from top to bottom) at $t=0$, $t\approx1.45/ H_*$, $t\approx7.23/H_*$, $t\approx 10.13/ H_*$. Case A shows the existence of an apparent horizon (the outer-most trapped surface) around $r\approx 0.25/H_*$ in on the bottom panel.
    } 
    \label{fig:ah_example}
\end{figure}

In order to unambiguously show that a black hole is formed, we need to evaluate the expansion, \Eq{eq:expansion}, and demonstrate the existence of a marginally trapped surface. \Fig{fig:ah_example} demonstrates the existence of a trapped surface for case A, and the absence of a trapped surface for case B.  For case A the apparent horizon forms between $t=7.2/H_\star$ and $t=10.1/H_\star$.  For all simulations with larger initial compaction than case A, $C(R_c)>0.85$, horizons form at earlier coordinate time, $t$.

\section{Conclusions}
\label{sec:discussion}

PBHs can form as a result of a variety of mechanisms in the early Universe, including the direct collapse of over-dense regions resulting from quantum fluctuations that froze out during inflation. 
In this work, we investigate the formation of black holes via the collapse of over-dense regions in a radiation-dominated universe. 
Specifically, we perform numerical simulations of a radiation fluid in fully nonlinear general relativity with spherically symmetric Gaussian over-densities as initial data. 
As opposed to setting the boundary conditions to match FLRW dynamics outside of the computational region by using the pressure gradients at the outer boundary of the box or enforcing an expansion history at the edge of the boundary, we employ periodic boundary conditions on a Cartesian cube, and allow the matter content to drive the dynamics of the simulation without explicitly enforcing asymptotic FLRW cosmology on the boundary.  

We parameterize our initial data by $\zeta$ and directly solve the Hamiltonian constraint, as an alternative to employing the gradient expansion \cite{Shibata:1999zs,Harada:2015yda,Salopek:1990jq}.  As discussed in \cite{Baumgarte:2025vvs}, balancing the extrinsic and intrinsic curvature for initial data lead to strong-field and weak-field branch solutions. We choose initial conditions that spans both branches.

We calculate a critical threshold of $\delta_c \approx 0.8$; over-dense regions whose cosmological horizon-crossing compactness is above this level form PBHs. This result is generally in agreement with current analytical estimates of the critical threshold which find $0.7 \leq \delta_c \leq 1.15$ 
\cite{Musco:2020jjb}. 
Additionally, our results are consistent with previous numerical studies of PBH formation \cite{Yoo:2020lmg,Yuwen:2026hxu}; in particular, we agree with the results presented by Yoo {\sl et al.} \cite{Yoo:2020lmg}, who investigated spherical initial profiles and expressed the collapse threshold in terms of the amplitude $\mu$ of the curvature perturbation $\zeta$, finding that black hole formation occurs for $\mu \gtrsim 0.805$. This value is in agreement with the amplitude of our case A, providing an independent numerical validation of their threshold. 
This suggests that our simulations accurately capture the onset of gravitational collapse and are consistent with the current understanding of PBH formation thresholds.

In addition to validating and extending previous work, our results give us confidence that our code accurately simulates PBH formation in a radiation-dominated cosmology.  
This paves the way for more complex and realistic initial energy density configurations beyond spherical symmetry, such as multiple modes, to be investigated in future work, as well as studies of the subsequent evolution and accretion onto PBHs that may or may not form.

\acknowledgments
We thank Josu Aurrekoetxea, Sam Brady, Eve Currens, Ericka Florio, David Kaiser, Alan Guth, and Chul-Moon Yoo for useful discussions.  

This work is supported in part by National Science Foundation grant PHY-2308821 to Bowdoin College.  M.G.~is supported by the U.S.~Department of Energy, Office of Science, Office of Advanced Scientific Computing Research, Department of Energy Computational Science Graduate Fellowship under Award Number DE-SC0023112. J.T.G.~and A.M.~are supported in part by the National Science Foundation, PHY-2309919, awarded to Kenyon College.  

This report was prepared as an account of work sponsored by an agency of the United States Government. Neither the United States Government nor any agency thereof, nor any of their employees, makes any warranty, express or implied, or assumes any legal liability or responsibility for the accuracy, completeness, or usefulness of any information, apparatus, product, or process disclosed, or represents that its use would not infringe privately owned rights. Reference herein to any specific commercial product, process, or service by trade name, trademark, manufacturer, or otherwise does not necessarily constitute or imply its endorsement, recommendation, or favoring by the United States Government or any agency thereof. The views and opinions of authors expressed herein do not necessarily state or reflect those of the United States Government or any agency thereof.


\appendix 

\section{Relativistic Fluids in the 3+1 Decomposition}
\label{sec:appendixB}

In this appendix, we describe how we embed a relativistic fluid into our simulations.  In \Sec{subsec:definedynamical} we define dynamical degrees of freedom and show how they arise from the primitive fluid variables.  In \Sec{subsec:programunits}, we introduce the dimensionless variables we employ in our numerical implementation.

\subsection{Source Tensor and Scalar}
\label{subsec:definedynamical}

The stress energy tensor for a relativistic, perfect fluid with constant equation of state parameter $w = 1/3$, i.e.~a radiation fluid with $p=\rho_0/3$, is given by 
\begin{equation}
    T^{\mu \nu} = \frac{4}{3} \rho_0 U^{\mu} U^{\nu} + \frac{\rho_0}{3} g^{\mu \nu},
\end{equation}
where $\rho_0$ is the rest-frame energy density of the fluid (where we use the subscript zero to distinguish $\rho_0$ from the ADM energy density $\rho$ defined in \Eq{eq:defofrho} below).  These variables, $\rho_0$ and $U^\mu$, are the so-called {\sl primitive variables} (see, e.g., \cite{Baumgarte:2010ndz}) which fully characterize the dynamics of the fluid system; however, it is not guaranteed that these evolution variables are well suited for 3+1 evolution. Therefore, we define a set of dynamical variables, built from the primitive variables, as follows.
    
The source-terms in the BSSN formalism are
\begin{align}
    \rho &= n_a n_b T^{ab}, \label{eq:defofrho}\\
    S_i &= - \gamma_{i\mu} n_{\nu} T^{\mu\nu}, \label{eq:vectorsource}\\
    S_{ij} &= \gamma_{i\mu}\gamma_{j\nu} T^{\mu\nu}, \label{eq:tensorsource}
\end{align}
which are defined as the projections of the stress-energy tensor onto our 3+1 decomposition.  The spatial hypersurfaces are defined by the time-like normal vector, 
\begin{equation}
n_\alpha = (-\alpha, 0,0,0),
\end{equation}
which also defines the spatial projection tensor,
\begin{equation}
\gamma_{\mu\nu} = g_{\mu\nu} - n_\mu n_\nu.
\end{equation}
The total mass-energy density measured by a normal observer is given by $\rho$, the momentum density is $S_i$, and the stress is $S_{ij}$, where the trace of the stress is $S = \gamma^{ij} S_{ij}$. 

It is also useful to also define the 3-velocity of our fluid on our spatial hypersurfaces, $v^i$, as
\begin{equation}
\label{eq:defofv}
    v^i = \frac{1}{W}{\gamma^{i}}_\alpha U^\alpha = \frac{U^i}{W}+\frac{\beta^i}{\alpha},
\end{equation}
where $W$ is the Lorentz factor between normal and fluid observers, $W =\alpha U^0$.  

The $00$-component of the stress-energy tensor is associated with the energy of the fluid,
\begin{equation}
\label{eq:defT00}
T^{00} = \frac{4}{3} \rho_0 U^{0} U^{0} + \frac{\rho_0}{3} g^{00} = \frac{\rho_0}{\alpha^2}\left(\frac{4}{3}W^2-\frac{1}{3}\right),
\end{equation}
which gives us a natural definition for the  fluid energy density, $\mathcal{E}$,
\begin{equation}
    \label{eqn:curlyEdef}
\mathcal{E} \equiv \alpha T^{00} = \frac{\rho_0}{\alpha} \left(\frac{4}{3}W^2-\frac{1}{3}\right).
\end{equation}
The energy density, \Eq{eq:defofrho}, is  
\begin{align}
\rho &= n_\alpha n_\beta T^{\alpha \beta} = \alpha \mathcal{E}.
\end{align}

To define the momentum density $S_i$, we start with the the $0i$-components of the stress energy tensor, 
\begin{equation}
    T^{0i} = \frac{4}{3} \rho_0 U^{0} U^{i} + \frac{\rho_0}{3} g^{0i} = \frac{\rho_0}{\alpha} \left(\frac{4}{3}WU^i +\frac{1}{3}\frac{\beta^i}{\alpha}\right),
\end{equation}
along with \Eq{eq:defofv} to define the fluid momentum density, $\mathcal{P}^i$, as 
\begin{equation}
\label{eqn:curlyPdef}
\mathcal{P}^i \equiv \gamma^{ij} \frac{4}{3}W^2 \rho_0 \, v_j,
\end{equation}
so that 
\begin{equation}
\label{eq:defT0i}
T^{0i} = \frac{1}{\alpha}\mathcal{P}^i - \mathcal{E}\frac{\beta^i}{\alpha}.
\end{equation} 

With these definitions, \Eqs{eqn:curlyEdef}{eqn:curlyPdef}, the source vector, \Eq{eq:vectorsource}, is 
\begin{equation} 
S_i = -\gamma_{i\mu} n_\nu T^{\mu\nu} =  \gamma_{ij}\mathcal{P}^j,
\end{equation}
and the source tensor, \Eq{eq:tensorsource}, becomes
\begin{equation}
\label{eqn:Sij}
   S_{ij} = \frac{\mathcal{P}_i\mathcal{P}_j}{\alpha \mathcal{E} + \rho_0/3 }
+ \frac{\rho_0}{3}\gamma_{ij}.
\end{equation} 

In practice, the evolution equations for the fluid (see \App{sec:appendixC}) can be written more compactly in terms of 
\begin{align}
    E &= \sqrt{\gamma} \alpha \mathcal{E} \\
    P^i &= \sqrt{\gamma}\mathcal{P}^i.
\end{align}
The equations of motion also require us to calculate derivatives of $S^{ij}$; therefore, we defined an additional tensor
\begin{equation}
\label{eq:sijcode}
    S^{ij}_{\rm v} =  \frac{\alpha P^iP^j}{E + \sqrt{\gamma}\rho_0/3}  + \frac{\alpha \rho_0}{3}\gamma^{1/6}\bar{\gamma}^{ij}.
\end{equation}
From the dynamical variables $E$ and $P^i$, we recover the rest-frame density, $\rho_0$, on each slice; to do this, we first calculate $\mathcal{P}_i\mathcal{P}^i$ and substitute the identity $W^{-2} = 1-v_iv^i$, 
\begin{equation}
\mathcal{P}_i\mathcal{P}^i =  \alpha^2 \mathcal{E}^2 -2 \alpha \mathcal{E}\frac{\rho_0}{3} - \frac{\rho_0^2}{3}.
\end{equation}
We solve this quadratic equation to find 
\begin{align}
 \rho_0 = -\alpha \mathcal{E} + \sqrt{4\alpha^2 \mathcal{E}^2 - 3\mathcal{P}_i\mathcal{P}^i},
\end{align}
where we keep the positive square root.

In summary, the gravitational sources as well as the rest-mass density can be written in terms of our derived dynamical quantities, $E$ and $P^i$,
\begin{align}
        \rho &= \frac{E}{\sqrt{\gamma}}\\
        S_i &= \bar{\gamma}_{ij} \gamma^{-1/6} P^j\\
        S_{ij} &= \frac{1}{\alpha \sqrt{\gamma}}S_{ij}^{\rm v} =\frac{ \gamma^{1/6}}{\alpha} \bar{\gamma}_{ik}\bar{\gamma}_{jl}S^{kl}_{\rm v}\\
        \rho_0 & =  -\frac{E}{\sqrt{\gamma}}+ \sqrt{4\frac{E^2}{\gamma}  - 3\frac{1}{\gamma} P^i P_i}.
\end{align}

\subsection{Dimensionless Units}
\label{subsec:programunits}
As with all numerical implementations, it is necessary to define dimensionless versions of our dynamical degrees of freedom.  Here we use the subscript ``pr" to refer to these {\sl program} variables used in our code.  

We start by defining dimensionless spacetime intervals,
\begin{equation}
dx_{\rm pr}^\mu = B\,dx^\mu,
\end{equation}
where $B$ carries units of mass. Since energy density has units of mass to the fourth power, we define a dimensionless energy density according to   
\begin{equation}
\rho_{\rm pr}  = \frac{\rho}{B^2m_{\rm pl}^2 }.
\end{equation}
In practice, we set $B$ such that
\begin{equation}
\label{eq:rhorescale}
     \bar{\rho}_{*,\rm pr}  = \frac{\bar{\rho}_*}{B^2m_{\rm pl}^2 } = 1,
\end{equation}
where $\bar{\rho}_*$ is the initial, asymptotic energy density as discussed in \Sec{sect:initialconditions}.  
The dynamical variables, which also carry units of mass to the fourth power, require the same rescaling as \Eq{eq:rhorescale},
\begin{align}
    E_{\rm pr} &= \frac{E} {B^2 m_{\rm pl}^2} \\
    P^i_{\rm pr} &= \frac{P^i }{B^2 m_{\rm pl}^2}.
\end{align}
The 4-velocity, $U^\mu$, and the 3-velocity, $v^i$, are already dimensionless.  

\section{Fluid Evolution Equations}
\label{sec:appendixC}

In this appendix, we derive the evolution equations for our derived fluid variables, $E$ and $P^i$, as defined in \App{sec:appendixB}.

We adopt a version of the Wilson scheme,\footnote{As noted in \Sec{sec:numerics}, we do not anticipate that shocks will develop prior to black hole formation; one could also use the Valencia formulation \cite{Banyuls:1997zz,Montero:2013pca}, which, as a flux-conservative formulation, is better suited for the implementation of shock capturing schemes.} a scheme for formulating the fluid variables which is often used to incorporate fluids into fully nonlinear simulations  \cite{1972ApJ...173..431W,1979sgrr.work..423W,1984ApJ...277..296H} (see, e.g., \cite{Baumgarte:2010ndz} for a general overview). 

To derive the time-evolution of the energy density, we take the divergence of the contraction of the stress energy tensor with the normal vector, $n_\alpha T^{\alpha \beta}$, 
\begin{equation}
\label{eq:firststepwilson}
\nabla_\beta \left(n_\alpha T^{\alpha \beta}\right) = \frac{1}{\sqrt{-g}}\partial_\beta \left(\sqrt{-g} n_\alpha T^{\alpha \beta}\right). 
\end{equation}
Local conservation of stress energy also requires that 
\begin{align}
\label{eq:secondstepwilson}
\nabla_\beta \left(n_\alpha T^{\alpha \beta}\right) 
&= T^{\alpha \beta}\nabla_\beta n_\alpha.
\end{align}
Combining \Eq{eq:firststepwilson} with  \Eq{eq:secondstepwilson} yields
\begin{align}
\label{eq:fulleomnotsimplified}
\frac{1}{\sqrt{-g}}\,
\partial_\beta \!\left(\sqrt{-g}\, n_\alpha T^{\alpha\beta}\right)
&= T^{\alpha\beta}\nabla_\beta n_\alpha.
\end{align}
\Eq{eq:fulleomnotsimplified} contains only the equation of motion for the energy density of the fluid, which can be expressed in terms of the components of the stress energy tensor, 
\begin{widetext}
\begin{equation}
\partial_0 \left(\alpha^2\sqrt{\gamma} \,T^{00}\right)
+ \partial_i \left(\alpha^2\sqrt{\gamma}\, T^{0i}\right)
= \alpha\sqrt{\gamma}\, T^{00}\, \partial_0\alpha 
  + \alpha\sqrt{\gamma}\, T^{0i}\, \partial_i\alpha  
 - \alpha\sqrt{\gamma}\, T^{\alpha\beta}
   {\Gamma^0}_{\alpha\beta}\, \alpha ,
\end{equation}
or, written in terms of the dynamical variables, $E$ and $P^i$,
\begin{equation}
\label{eqn:Eom_E}
\partial_0 E
+ \alpha\, (\partial_i P^i)
 - \beta^i (\partial_i E)
 - E\, (\partial_i \beta^i)
= -2 P^i (\partial_i \alpha)
   + P^i \beta^j K_{ij}
   + K_{ij} S^{ij}_{\rm v}
   - P^j \beta^i K_{ij},
\end{equation}
where we found \cite{Golovnev:2013fj} to be a great resource for calculating the Christoffel coefficients.
\end{widetext}

The equations of motion for the momentum density $P^i$ begin similarly by considering the conservation of local energy-momentum $\nabla_\alpha {T^{\alpha}}_\beta = 0$ \cite{1972ApJ...173..431W,1979sgrr.work..423W,1984ApJ...277..296H,Baumgarte:2010ndz}. These four equations take the form
\begin{equation}
\label{eqn:wilson}
\partial_0 \left(\sqrt{-g} {T^0}_\mu \right) + \partial_i \left(\sqrt{-g} {T^i}_\mu \right) = \frac{1}{2}\sqrt{-g}T^{\alpha \beta}\partial_\mu g_{\alpha \beta}.
\end{equation}
Employing \Eqs{eq:defT00}{eq:defT0i}, and choosing the spatial, $\mu = j$, terms of \Eq{eqn:wilson}, we arrive at the dynamical equations for $P^i$.  These can be written in terms of the components of the stress-energy tensor,
	\begin{equation}
    \label{eqn:peom_intermediate}
		\partial_t \left(\sqrt{-g} {T^0}_j \right) + \partial_i \left(\sqrt{-g} {T^i}_j \right) = \frac{1}{2}\sqrt{-g}T^{\alpha \beta}\partial_j g_{\alpha \beta}.
	\end{equation}
or in terms of the dynamical variables,
\begin{widetext}
\begin{equation}
\partial_0 P_j
= -\gamma_{kj}\, \partial_i S^{ik}_{\rm v}
 - S^{ik}_{\rm v}\, \partial_i \gamma_{kj}
 + P_j\, \partial_i \beta^i
 + \beta^i\, \partial_i P_j -E\, \partial_j \alpha
 + \frac{1}{2} P^i \beta^k\, \partial_j \gamma_{ik}
 + P_k\, \partial_j \beta^k  + \frac{1}{2} S^{ik}_{\rm v}\, \partial_j \gamma_{ik}
 - \frac{1}{2} P^k \beta^i\, \partial_j \gamma_{ik} \label{eqn:Eom_P} .
\end{equation}
\end{widetext}
\Eqs{eqn:Eom_E}{eqn:Eom_P} provide a closed system that completely describes the evolution of the fluid within the 3+1 decomposition.


\bibliography{bibliography}

\end{document}